\documentclass[12pt,preprint]{aastex}

\newcommand{\be}{\begin{eqnarray}}
\newcommand{\ee}{\end{eqnarray}}
\newcommand{\Msun}{\mbox{M$_{\odot}$}}

\newcommand {\tauff} {\tau_{\scriptsize{\mbox{ff}}}} 
\newcommand {\epsff} {\epsilon_{\scriptsize{\mbox{ff}}}}

\shorttitle{Evolutionary tracks of young massive star clusters}
\shortauthors{Pfalzner et al.}

\begin{document}


\title{The evolutionary tracks of young massive star clusters}


\author{S. Pfalzner\altaffilmark{1}, G. Parmentier\altaffilmark{2}, M. Steinhausen\altaffilmark{1}, K. Vincke\altaffilmark{1}, and K. Menten\altaffilmark{1}}
\affil{$^1$Max-Planck-Institut f\"ur Radioastronomie, Auf dem H\"ugel 69, 53121 Bonn, Germany}
\affil{$^2$Astronomisches Rechen-Institut, M\"onchhofstr. 12-14, 69120 Heidelberg, Germany}
\email{spfalzner@mpifr.de}




\begin{abstract}
Stars mostly form in groups consisting of a few dozen to several ten thousand members.  For 30 years, theoretical models provide a basic concept of how such star clusters form and develop: they originate from the gas and dust of collapsing molecular clouds. The conversion from gas to stars being incomplete, the left over gas is expelled, leading to cluster expansion and stars becoming unbound. Observationally, a direct confirmation of this process has proved elusive, which is attributed to the diversity of the properties of forming clusters. Here we take into account that the true cluster masses and sizes are masked, initially by the surface density of the background and later by the still present unbound stars. Based on the recent observational finding that in a given star-forming region the star formation efficiency depends on the local density of the gas, we use an analytical approach combined with \mbox{N-body simulations, to reveal} evolutionary tracks for young massive clusters covering the first 10 Myr.  Just like the Hertzsprung-Russell diagram is a measure for the evolution of stars, these tracks provide equivalent information for clusters. Like stars, massive clusters form and develop faster than their lower-mass counterparts, explaining why so few massive cluster progenitors are found. 
\end{abstract}




\section{Introduction}

Massive star clusters\footnote{Here we use the term 'cluster' for any stellar group that forms spatially correlated regardless of whether it is bound or unbound} containing tens of thousands of stars form from the high-density parts of collapsing giant molecular clouds (GMCs).  In our study we concentrate on the formation and dynamical evolution of such clusters in the solar neighborhood\footnote{We exclude the compact massive clusters that are nearly exclusively found close to the Galactic center and the spiral arms where possibly much higher star formation efficiencies are achieved.} (for a list of such clusters see Wolff et al. 2007). Here observations find that massive clusters preferentially form where GMC filaments intersect and can be roughly approximated by centrally condensed spherical structures (Myers 2011, Andre et al. 2013).   

 During the last decade, there has been considerable observational progress in understanding the global, cluster-averaged properties in the solar neighborhood(for example, Evans et al. 2009,  Guarcello et al. 2010,  Townsley et al. 2011, and references therein, Megeath et al. 2012, Feigelson et al. 2013): 

\begin{itemize}
\item There seems to exist a relation between mass and size (red circles in Fig.1) for clusters still embedded in their natal gas. At face value, this implies that the cluster mass determines the cluster radius (Adams et al. 2006).  Alternatively, it might be interpreted as a developmental sequence, as more massive clusters have to build up somehow over time (Pfalzner 2011). 
\item Comparing the masses of the stellar component, $M_{s}$, to the gas masses, $M_{g}$, in a variety of star-forming clusters  it was found that typically only 10\%--35\%  of the gas is transformed into stars (Lada  et al. 2010). 
\end{itemize}

The reaction of the star cluster to the gas loss which ends star formation is highly deterministic and has been studied in numerous works (for example, Hills 1980, Adams 2001, Baumgardt \& Kroupa 2007, K\"upper et al. 2008, Parmentier \& Baumgardt 2012). Recently, Pfalzner \& Kaczmarek (2013a) showed that the observed mass and radius development after gas expulsion of the most massive clusters in the solar neighborhood corresponds to the predicted behavior of systems having formed with a 30\% star formation efficiency (SFE).  The SFE is defined as $ \epsilon_{SFE} = M_{s}/(M_s+M_{g})$.

Most previous investigations implicitly assumed that the SFE might differ in the various star-forming regions but is constant throughout a single cluster-forming volume.
However, recent {\it spatially resolved} observations (Gutermuth et al. 2011, Ybarra et al. 2013) of star forming regions show that {\it locally} the SFE depends on the local gas density. Mostly, they find that approximately $\Sigma_{s}\propto \Sigma_{gas}^2$, where $\Sigma_s$ and $\Sigma_{gas}$ are the stellar and gas surface density, respectively\footnote{Note, there are some clusters that have shallower or steeper dependencies, which might be reflecting that they are either at the beginning of the star formation process or already in the gas expulsion stage.}. The {\it local} SFE being a function of the gas density means that in areas of high gas density the same amount of gas is transformed into more stars than in regions of low gas density.  

Here we apply this concept of a local SFE self-consistently throughout the cluster-formation and expansion process and compare the results with observed cluster properties (Lada \& Lada 2003, Wolff et al. 2007). This requires taking into account that the true cluster masses and sizes are masked, initially by the surface density of the background and later by the still present unbound stars \citep{parmentier:13,pfalzner:13a}. In the past it was assumed that  the properties of forming clusters are very diverse. However, this apparent diversity can be largely explained as an artefact caused by the  variation in evolutionary rates in cluster development caused by the differences in gas density. 
The here presented method reveals mass-dependent evolutionary tracks for young massive clusters for the first 10 Myr of their development.

\section{Method}

We apply two different models to describe, first, the star-formation phase and, second, the cluster expansion phase of massive clusters. However, despite involving two methods the treatment is self-consistent in the sense that 
both methods use the same dependence of the SFE on the local gas density. 
The parameter space covered by this study is summarized in Tables 1 and 2.

\subsection{Cluster formation phase}

Ab-initio calculations treating the detailed hydrodynamics of the gas, the dynamics of the stars and feedback processes (for an overview see Krumholz et al. 2014), are currently
not able to model the formation of massive clusters ($M_s>$ 10 000 \Msun) due to computational constraints.  
Although feedback processes are definitely important for the specific development of individual clusters, a semi-analytical model (Parmentier \& Pfalzner 2013) can reproduce the observed local density dependence (Gutermuth et al. 2011), the mass-size relation of clusters (Adams et al. 2006) and the typical averaged SFE (Lada et al. 2010).  Here we only give a brief outline of the model, details can be found in Parmentier \& Pfalzner (2013). 

This model assumes that the SFE per free-fall time $\epsff$ is constant,  where the free-fall time, $\tauff$, scales with the gas volume density, $\rho_{g}$, as $ \tauff \propto \rho_g^{-1/2}.$ This means that every  $\tauff$ a fraction  $\epsff$ of the local gas mass is converted into stellar mass. The star cluster forming sub-units of giant molecular clouds, in the following referred to as clumps, are represented by spheres with a radial density gradient.  The radial density gradient of the gas results in the star formation being fastest in the dense clump center and slower at larger distances. Note that although formally similar, this model differs from that of Krumholz et al. (2005, 2007) as they define $\tauff$ globally for an entire cloud, whereas here all parameters are defined locally in the clump. 

The star formation process can be described by relatively simple differential equations for the temporal development of the gas and stellar density, $\rho_{g}(r,t)$ and $\rho_{s}(r,t)$, (see Parmentier \& Pfalzner 2013).
The solution for the stellar density is 
  \be 
\rho_s(r, t) & = &  \rho_0 (r) -\left(\rho_0(r)^{-1/2} +\sqrt{ \frac{8G}{3\pi}} \epsff t \right) ^{-2} \ee
where  $\rho_0(r) = \rho_0(r, t=0)$ is the initial clump density profile. Here we adopt an isothermal sphere given as $\rho_0(r) = (M_0)/(4 \pi R_g r^2)$ with $R_g$ being the clump radius.
The integration over the clump volume  provides the total stellar mass (hence the global SFE) 
and the half-mass radius of the cluster.

When comparing the results with the observed cluster properties in the formation phase one has to take into account that the observational data are surface density limited. This means that 
the “wings” of the stellar component are concealed, so that the observed masses and radii are smaller than the actual ones
(see fig.~5 in Parmentier \& Pfalzner 2013). Here we assumed a surface density limit of  $\Sigma_{bg}$= 40 \Msun pc$^{-2}$ (Carpenter et al 2000).  
In Parmentier \& Pfalzner 2013) it was demonstrated that the mass-radius relation for embedded clusters in the Lada \& Lada (2003) sample can be reproduced very well with this approach assuming an initial clump mass \mbox{$M_0$= 10\,000 \Msun}. 
 Obviously the surface density limit varies depending on the method used to select cluster members, the Galactic coordinates of the cluster, and its distance.  As a consequence, the data taken in different star forming regions and/or with different methods might shift slightly in the mass-radius plane but the slope, the here relevant parameter, remains basically the same as long as the cluster is surface-density limited (Pfalzner et al., in preparation).

\subsection{Cluster expansion phase}

When the star cluster has formed, the gas is expelled by various feedback processes. 
For massive clusters this gas expulsion is very fast ($<$ 1 Myr) due to the presence of many \mbox{O stars} (for example, Goodwin 1997; Melioli \& de Gouveia dal Pino 2006,  Lopez et. al. 2011, Moeckel et al. 2012, Plunkett et al. 2013, Krumholz et al. 2014, and references therein). In this case  instantaneous gas expulsion can be assumed and the cluster modeled as a stellar system out of virial equilibrium. For the low-mass clusters we performed also simulations modeling the gas expulsion as a decreasing background potential. For the investigated cases we find that as long as the gas is expelled in less than 1 Myr, the results show only a slight increase in bound mass and half-mass radius at age of 20 Myr compared to those with instantaneous gas expulsion. These changes are so small that they would not be detectable in current observations. Therefore we conclude that modeling the gas expulsion as instantaneous is definitely justified for the massive clusters ($>$ 5000 \Msun), whereas for the lower-mass clusters there exists some uncertainty concerning this point. 

In the standard method a parameter $\epsilon$ is defined, corresponding to the SFE, which describes how much the cluster deviates from virial equilibrium after gas expulsion. This value corresponds to a constant SFE throughout the star forming clump. Here the fundamental difference to previous work is that, we define not a global, but a {\it local} parameter $\epsilon (r, t)$ that directly corresponds to  the local, radial varying SFE defined in Section 2.1. 

Observations show that the stellar density profiles of young clusters
just before gas expulsion are best represented by King models with $W_0 >$ 7 (for example, Hillenbrand \& Hartmann 1998).  Therefore, the stars are distributed so that the stellar density follows a $W_0$ = 9 King profile. The velocities of the stars are chosen in such a way as to represent the departure from equlibrium after gas expulsion and at the same time take the radial dependence of $\epsff$ at the end of the star formation process into account. The masses of the stars are drawn from the initial mass function as observed in young clusters (Kroupa 2001) as otherwise the effect of encounters are severely under-represented (Pfalzner \& Kaczmarek 2013a).  
Stellar masses span from the hydrogen burning limit \mbox{($m_{\mathrm{star}} < 0.08$ $\Msun$)} up to \mbox{$m_{\mathrm{star}} = 150$ $\Msun$}.

All stars are modeled as initially single,  which is a standard procedure that significantly reduces the computation time. The few binaries that form during the simulation are treated as such in the remainder of the simulation. It has been demonstrated that including binaries leads to slight additional cluster expansion (Kaczmarek 2012). However, this additional expansion is of the order of 10\% of the half-mass radius and therefore smaller than the observational uncertainties.  
The clusters are modeled  as initially non-mass segregated, as it is an open question whether the often observed mass segregation in young massive clusters is the result of dynamical evolution or primordial.  

We model the reaction of the cluster to gas expulsion by direct N-body methods using the code NBODY6-GPU (Aarseth 2003, Nitadori \& Aarseth 2012). Here no tidal field is included in the simulations. The campaign covered simulations with  1\,000, 3\,000, 6\,000, 12\,000, and 30\,000 stars,  these correspond to models A, B, C, D, and E in Table 2, respectively. The simulations
were performed for each parameter set repeatedly to obtain statistically robust results (see \mbox{Table 2}).
We choose the initial half-mass radius as $r_{hm}$=1.3 pc as a previous parameter study, assuming a spatially constant SFE, revealed that a half-mass radius at between 1 pc and 3 pc gives the best fit to the observed massive cluster mass-radius development with cluster age (Pfalzner \& Kaczmarek 2013b).

When the stars become unbound after gas expulsion, they need some time (5-10 Myr) to leave the cluster. Unless high-quality proper motion data are available, observations would still regard these now unbound stars as cluster members during that timespan. In our diagnostics we take this observational fact into account by initially considering all stars, whether bound or not, within a predefined field of view of 40 pc x 40 pc (Bastian \& Goodwin 2006) to determine the cluster properties. At later stages, we only take into account the bound stars as the two populations occupy spatially distinct areas then. In the time interval in between we interpolate between the two values (for details, see Pfalzner \& Kaczmarek 2013b, especially their Fig.~2).  
Such a large field of view is required because the typical half-mass radius of the bound population of massive clusters is 15-20pc at ages $>10$ Myr. Smaller fields of view would miss the outer cluster areas and as a result underestimate the cluster masses and half-mass radii considerably in the later stages of the cluster development.

\section{Results}

In the following we apply the methods described in Sect. 2.1 and 2.2 in a self-consistent way. This means that in a first step the parameters in both sets of simulations are chosen in such a way that they are at the same time consistent with  the observed cluster properties and correspond to each other.  

\subsection{Most massive clusters}

Due to the observational limitations imposed by the surface density limit,
only for the most massive clusters can the entire expansion process be observed despite the cluster expansion and star loss. By contrast, lower mass clusters quickly drop below the detection limit (Lada \& Lada 2003, Pfalzner 2013a). 
For the most massive, fully formed clusters in the solar neighborhood masses of a few ten thousand solar masses and half-mass radii $R_{hm}$ of 5--7 pc are observed at an age of 2--3 Myr (Pfalzner 2009, Portegies Zwart et al. 2010; blue squares in Fig.~1).  At an age of 1 Myr, when the star formation process ended, these clusters had a likely half-mass radius of 1--3 pc  (Pfalzner \& Kaczmarek 2013a). 

The cluster forming gas clumps must have had larger masses than the cluster mass, $M_0^{gas} > M_s = 10^4 \,\Msun$, and the clump outer radii exceeded the cluster half-mass radii, $R_g > R_{hm}$ = 1--3 pc at 1 Myr. Therefore we assumed the gas clumps to have a radius of $R_g$=6 pc  and to form clusters with stellar masses of \mbox{1.8 $\times$ 10$^4$ \Msun.} 

In principle, for a given clump radius there are several paths for a typical final cluster mass to be reached. The clump mass determines the density distribution and how fast the global SFE increases for a given $\epsilon_{\scriptsize{\mbox{ff}}}$. This means one needs a longer time to build up  the same total stellar cluster mass with a lower clump mass than with a higher one. Although the same cluster mass can be obtained with different clump masses, the local and total star formation efficiency in the resulting clusters will differ. As a consequence, the clusters will expand in different ways.  

First we determined the SFE distribution for clumps with $M_0^{gas}$ = 
\mbox{60\,000 \Msun}, 80\,000 \Msun, 100\,000 \Msun, and 120\,000 \Msun\, when they reach a cluster mass of \mbox{18 000 \Msun} with our analytical model  (corresponding to models S1, S2, S3 and S4, respectively, as detailed in \mbox{Table 1). For each of these models 15 simulations were performed to obtain adequate statistics}.  In these
calculations the SFE per free fall time is assumed to be $\epsff $=0.1. The sensitivity  to these parameters is discussed later in this section.  Afterwards we simulate the cluster development after gas expulsion  corresponding to these four different clump masses. 

Fig. 1 shows the simulations of the expansion phase  that lead to the best fit with the observations
(blue dashed line).  It corresponds to model S3 with a clump mass $M_0^{gas}$ = \mbox{100 000 \Msun}.
The mass-radius development in the formation phase of massive clusters is indicated in Fig. 1 by the top red solid line (half-mass radius). The maximum stellar mass is reached 0.7 Myr after star formation started; at this point the mean cluster age is $t_c$ = 0.45 Myr.  This star formation time corresponds approximately to one free-fall time of the cluster-forming clump. This is in agreement with the recent result  by  Dib et al. (2011).

The red dashed line shows the radius of total observed cluster area taking into account the surface density limit in observations. 
We see that there is very good agreement with the mass-radius development of embedded clusters in the  early phases, but a slight deviation when the cluster radius exceeds $ r_{hm} \approx$ 6 pc. The reason for the larger masses at larger cluster radii  in the mass-radius relation is that the observational data are dominated by the formation of star clusters from lower mass clumps. Low-mass clusters are more common and  they spent a longer time in this phase due to their longer formation time (see section 3.2).

Thus this simple model simultaneously  reproduces the observed relation between gas and stellar surface density in the cluster formation phase, and also the mass-radius relation. The reason is {\it not} that all the radiation and gas- and stellar-dynamics processes do not play a role, in the contrary, the relation of the gas to the stellar surface density is the end result of these processes. So if one
reproduces this relation one automatically obtains as well the cluster properties (see Fig.3 in Parmentier \& Pfalzner 2013). However, the real physical foundation of this relation can eventually only be found when detailed studies 
for such large systems will become available.

Figure 2 shows the cluster properties at the end of the formation phase. It can be seen that in the central cluster areas the gas density has dropped below the stellar density corresponding to a very high SFE ($>$ 80\%), whereas at the outskirts the local SFE is very low. The cluster averaged SFE is 18\%.
However, as observations usually concentrate on the inner cluster areas, one could expect a higher observed value of the order of 30\%.

In this model there are three free parameters: the free-fall time, $\tauff$, the time until gas expulsion ends the star formation process, $t_g$, and the radius of the initial gas clump, $R_{g}$. Naturally the results depend on the choice of these parameters. We performed a parameter study ($\epsff$ = 0.06, 0.1, 0.2) and found that varying $\epsff$ does not change the form of the mass-radius relation, just the time scale at which the different developmental stages are reached. The process is naturally limited to approximately 6 Myr, corresponding to a cluster age of roughly 3 Myr, the maximum age  observed for embedded clusters. One would need more precise observations of the end of the embedded phase as a function of the cluster mass to obtain tighter constraints on the development here.  Our simulations ($R_g$ = 3, 6, 10, 15 pc) show that the initial size of the gas clump has to lie in the range 3--12 pc as otherwise the clusters are either too small in size to obtain the observed expansion phase or  the mass of the most massive clusters cannot be reached before the gas becomes expelled.

\subsection{Lower mass clusters}

As for lower mass clusters a direct comparison with the observations during cluster expansion is not possible,
we assume that lower mass clusters develop in a similar fashion as  high-mass clusters.
 We performed the same types of calculations for lower clump masses  (models A-D), assuming that the gas expulsion happens always after one free-fall time, thereby implying the same global SFE.    As $t_g \propto \rho_g^{-1/2}$, these lower-mass clusters develop more slowly and consequently have a longer embedded phase (see Fig.~3). For lower clump masses, the same stellar mass or radius are reached at a later time. 

The different ages for clusters of the same mass and radius are explained as being in different developmental stages. The solid lines in Fig. 3 give the cluster masses as function of the time since the star formation process started. Naturally the cluster ages (dashed lines) are smaller than the time elapsed since star formation started as ages are time-averaged values per stellar mass increment. The lower gas density in low-mass clumps results in slower star formation which in return means a larger age spread in low-mass clusters than in high-mass ones (Parmentier et al., 2014). From our simulations we would expect an age spread of 0.7 Myr for our most massive clusters (model  A) and an age spread of 3.2 Myr for the low-mass clusters (model E). Indications for such an inverse correlation between cluster density and age spread have been found in recent observations (Reggiani et al. 2011, Kudryavtseva et al. 2012).

Figure 4 shows the relation between the cluster mass and radius for different initial clump masses. The colors indicate the age of the cluster at a given stage. The dashed line shows the observed cluster mass and radius, whereas the multicolor solid lines represents the values of the bound cluster population  for different initial clump masses.  The drop in the multicolor line indicates the large fraction of stars becoming unbound due to their kinetic energy exceeding the potential energy after gas expulsion. During a transition phase, that lasts approximately until 8 Myr for model A, the observed cluster mass exceeds considerably the bound cluster mass. Afterwards the bound and unbound stars are spatially distinct enough and the observed and bound cluster masses are the same. For lower mass clusters the time span where the observed cluster mass exceeds the bound cluster mass lasts longer. For instance, in model E it spans the entire 20 Myr studied here.

One result of our parameter study is that the cluster size at the moment of gas expulsion is for all clump masses approximately the same. This is a direct consequence of the assumption that the gas expulsion always happens at one free-fall time and that the gas clumps have the same size. Whether this is generally the case, should be tested by future hydrodynamical simulations. Figure~4 can be used as a map to determine the actual developmental stage of a given cluster, as the bound cluster mass can be determined from the observed mass, radius and age. 

The lower-mass clusters develop slower than the high-mass clusters, but will eventually reach sizes $>$ 10 pc only at ages $>$ 10 Myr. For the most massive clusters the unbound members have left the 40 pc x 40 pc field of view of the remnant cluster at 10 Myr. This is not the case for the lower mass clusters.

\section{Conclusion}

We describe here an evolutionary track model for the masses and sizes of extended massive clusters/associations over the first 10 Myr of their development. These tracks are based on the  assumption that the local SFE is a function of the gas density. This mass-radius relation for clusters is similar to the Hertzsprung-Russell diagram for stars. We find pre-defined tracks that the clusters follow
in their evolution as a function of the cluster mass. Like for stars, the more
massive clusters develop faster than low-mass clusters.

The obtained mass-radius relations are most secure for high-mass clusters. 
For lower mass clumps, and therefore clusters, such a direct comparison of the gas expulsion phase is 
currently not possible. The curves for the lower-mass clusters are
obtained assuming that the global SFE is the same as for high-mass clusters. However, in reality it might be  lower. In this case the tracks in Fig. 4 should be regarded as upper limits for low-mass cluster formation. Equally, the approximation of instantaneous gas expulsion is probably no longer justified for clusters with less than 1\,000 members (corresponding to $M_s \approx$ 600 \Msun).

As the mass change is related to a change in luminosity, and the "color" of the cluster alters with
age as more stars leave the pre-main sequence, 
the mass-radius tracks should translate into a luminosity-color relation.
Then there would be a direct correspondence to the HR diagram for stars. This has not been provided here as the result depends strongly on the chosen pre-main sequence models. Future work
should include a thorough investigation of the different options.

The hypotheses above can be tested in two ways: First, the slower formation of low-mass
clusters results in a larger age spread between the cluster stars. For example, a
cluster with a mass of 500 \Msun\  and a  formation time $t_g \propto M_0^{-1/2}$ have a six times larger age spread than a much more massive cluster with 20 000 \Msun (assuming both have the same initial size). First hints that this might be
the case are the small age spreads in the very dense, massive  clusters Westerlund 1 and NGC 3603 (Kudryavtseva et al. 2012).

Second, within this model, clusters of the same mass and radius can have
different ages as they form from clumps of smaller mass and therefore lower
density. This means that at a given mass/radius the older clusters
should have a smaller amount of residual gas as they develop into low-mass
clusters.

Some of the observational parameters taken as a guideline for these evolutionary tracks are not well determined. Here the actual ages are the most uncertain part, but relative ages should be well covered with this method. In summary, one needs tighter observational constraints to determine the relevant parameters of this model better. Especially better methods for determine absolute ages for young stars and mean ages for clusters would be needed to obtain a better calibration of the tracks. Another important improvement would come from a better discrimination between bound and unbound stars. This will likely be provided when the data from the GAIA mission become available and cluster membership can be better determined by photometric data. First observations of the $\gamma$ Velorum cluster show that such observations will be able to provide this distinction between bound and unbound members (Jeffries et al. 2014). For the embedded phase already recent data (for example, those provided in Kuhn et al. 2014) could be used to test the here presented model of cluster formation by comparing the stellar and gas components in the embedded systems. Hopefully this will result in larger samples of as well embedded clusters as well as clusters after gas expulsion and better determination of the actual development tracks.

\acknowledgments
We would like to thank the anonymous referee for the very constructive comments.

\clearpage



\begin{figure}
\epsscale{1.10}
\plotone{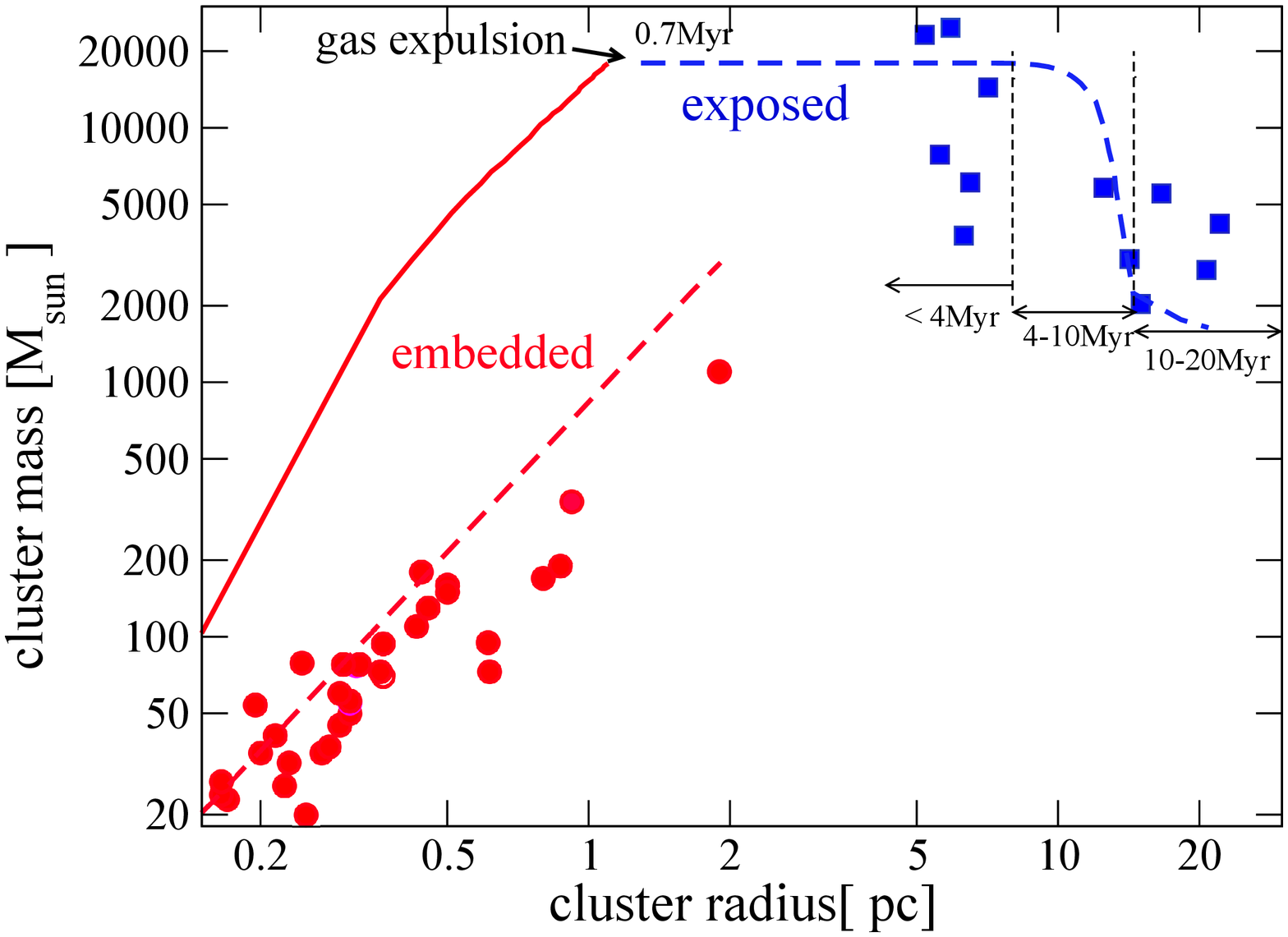}
\caption{Cluster mass as a function of radius. Attention should be given to the different definitions of the cluster radius for the different data sets in this plot. The radii of the observed clusters still forming stars (full red circles, data from Lada \& Lada 2003) are half the extent of the observed entire cluster sizes. The same definition is used in our star formation model for an initial clump of mass M= 100 000 \Msun\, taking into account surface density limitations in observation (dashed red  line). The red solid line shows how the half mass radius of the underlying cluster would develop. For the exposed clusters, where the star formation process is largely finished ( blue squares, data taken form Wolff et al 2007), the half-mass radius of these clusters is given. The blue dashed line shows the simulation result for our model cluster A. \label{fig1}}
\end{figure}

\begin{figure}
\epsscale{0.65}
\plotone{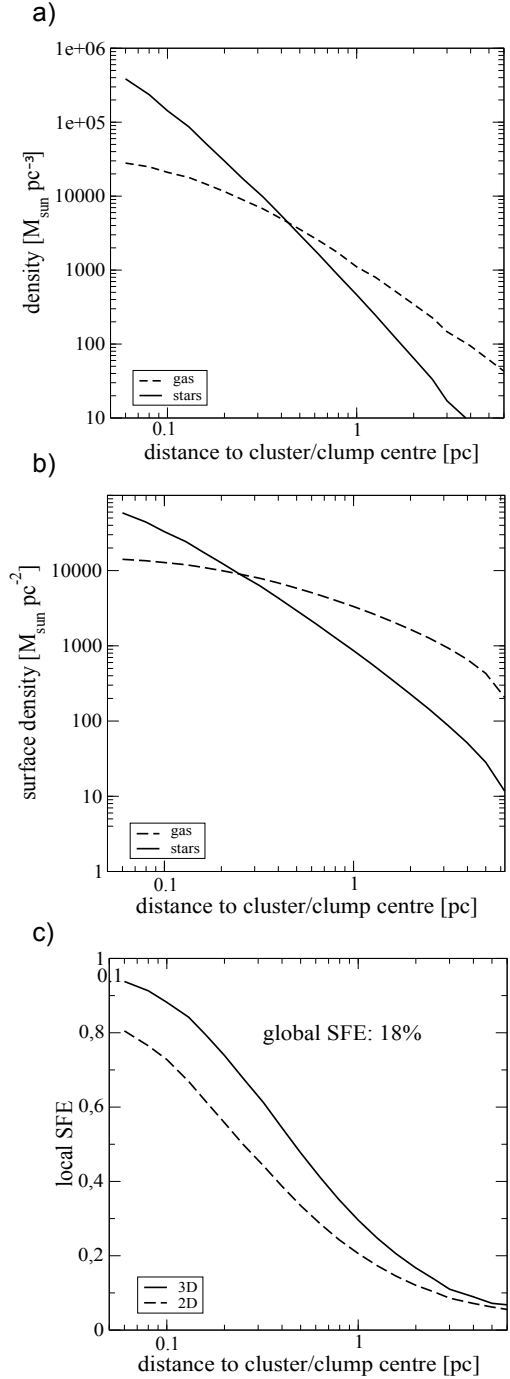}
\caption{Radial dependence of a) the gas (dashed line) and stellar density (solid line) at the moment of gas expulsion for the most massive clusters in the solar neighborhood (based on eqs. 19 and 20 of   Parmentier \& Pfalzner 2013). The corresponding surface densities are shown in panel b). Panel c) shows the resulting local SFE, here the projected observed 2-dimensional (dashed line) and calculated 3-dimensional values (solid line) are given. The integrated global SFE is indicated.\label{fig2}}
\end{figure}

\begin{figure}
\epsscale{0.9}
\plotone{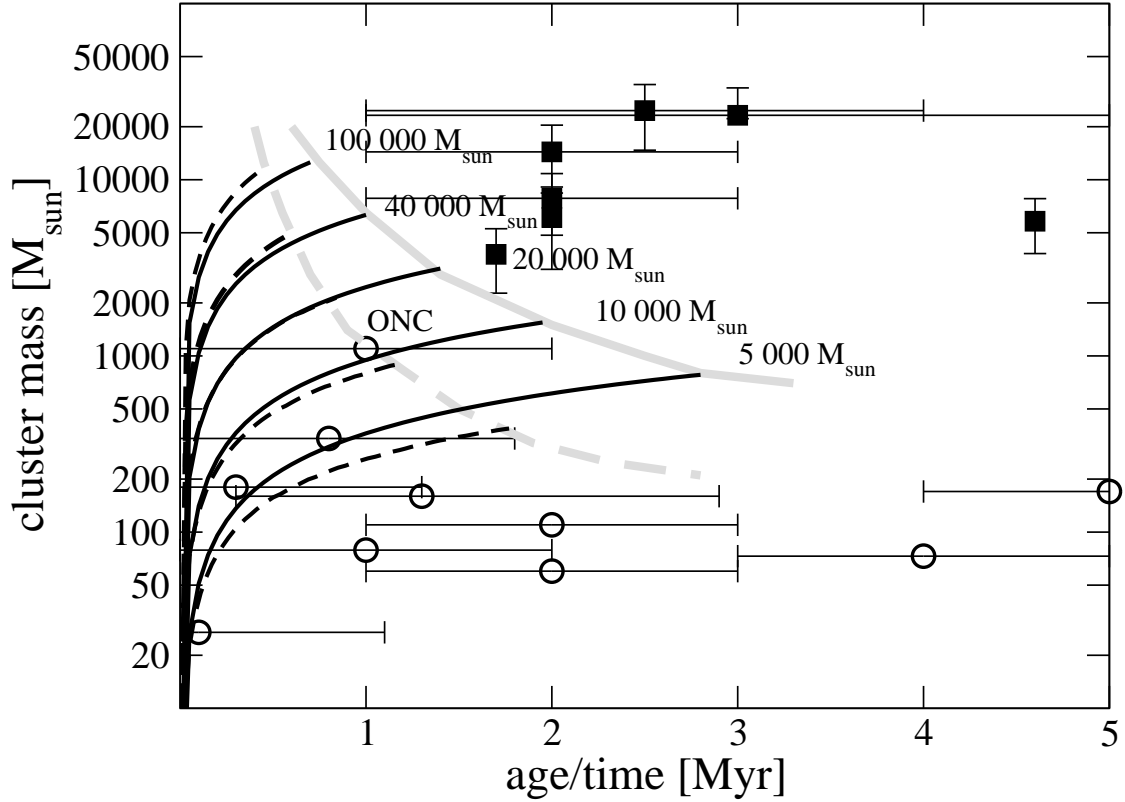}
\caption{Cluster mass as function of cluster age. The symbols show the observed values, circles depict embedded clusters, squares exposed clusters as in Fig.1. 
The solid lines show how the observed cluster mass increases as a function of time since star formation started  for clump masses of  5 000, 10 000, 20 000, 40 000, 100 000 \Msun and the dashed line the same as a function of cluster age. The grey line indicates the end of the star formation phase.\label{fig3}}
\end{figure}

\begin{figure}
\epsscale{.90}
\plotone{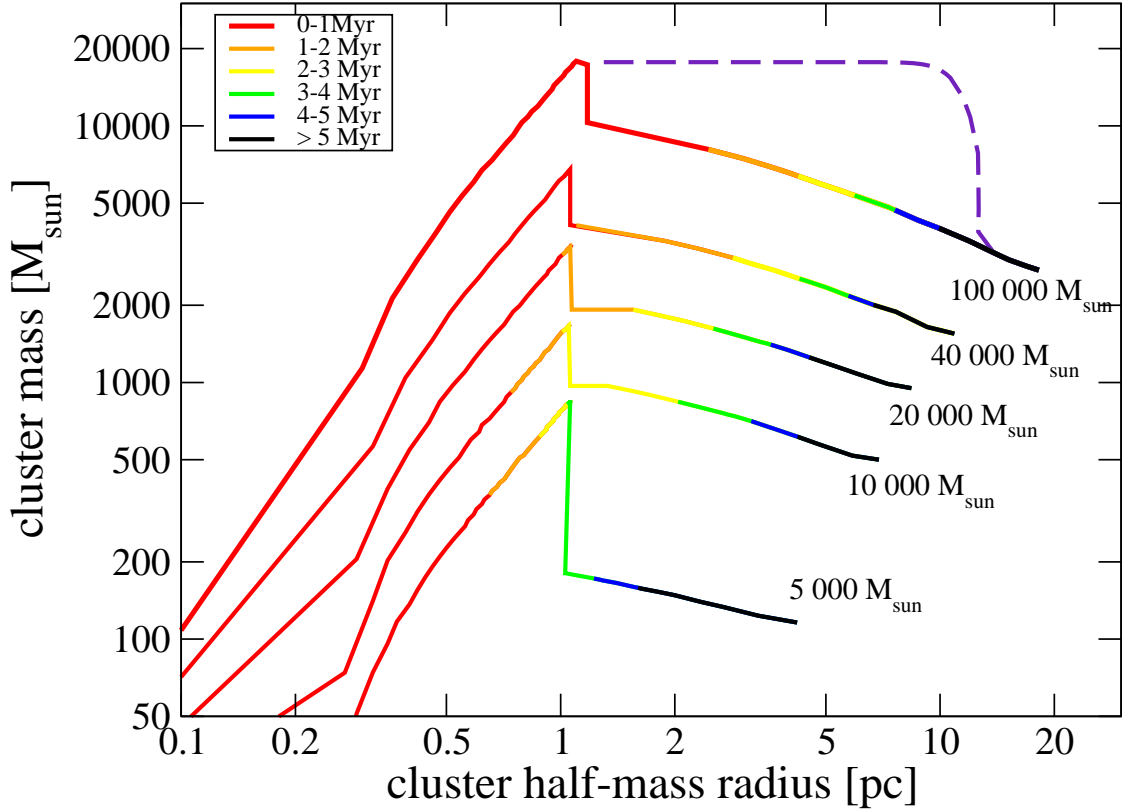}
\caption{The multi-color lines show the real cluster mass (bound stellar population) as a function of cluster half-mass radius for clump masses \mbox{100 000, 40 000, 20 000, 10 000 and, 5 000\Msun.} The colours indicate the cluster age. The purple  dashed line indicates the observed mass-radius development after gas expulsion of model A, corresponding to the top multi-colour line representative for a clump mass of 100\,000 \Msun. Note that this is the same line as indicated as blue dashed line in Fig.~1. \label{fig4}}
\end{figure}


\clearpage
\begin{table}
\begin{center}
\caption{Parameter of the clusters models used for the determination of a self-consistent representation in the formation and expansion phase\label{tbl-1}}
\begin{tabular}{lrrrr}
\tableline\tableline
Model & $M_0$ [\Msun]& $N$ & $M_s$ [\Msun] & No. of runs \\
\tableline
S1 & 60\,000     & 30\,000    &  17\,719    & 15\\
S2 & 80\,000    & 30\,000    & 17\,719   & 15\\ 
S3  = E& 100\,000   & 30\,000    & 17\,719   & 15\\ 
S4 & 120\,000   &30\,000 & 17\,719   &15\\ 
\tableline
\end{tabular}
\tablenotetext{a}{The first column gives the model name, $M_0$ depicts the initial clump mass, $N$ stands for the number of stars in the simulations, column 3 denotes the corresponding cluster mass, and the last column states the number of simulation runs performed for each set.}
\end{center}
\end{table}

\begin{table}
\begin{center}
\caption{Properties of Nbody simulations of clusters in expansion phase\label{tbl-1}}
\begin{tabular}{crrrr}
\tableline\tableline
Model & $M_0$ [\Msun]& $N$ & $M_s$ [\Msun] & No. of runs \\
\tableline
A & 5\,000     & 1\,000    &  293     & 400\\
B & 10\,000    & 3\,000    & 1\,759   & 200\\ 
C & 20\,000   & 6\,000    & 3\,513   & 100\\ 
D & 40\,000   &12\,000 & 7\,074   &50\\ 
E & 100\,000 &30\,000 & 17\,719 & 15\\
\tableline
\end{tabular}
\tablenotetext{a}{Same column definitions as in Table 1.  Note that for model A the value of $M_s/N$ is somewhat smaller than for the other cases as the lower number of cluster members means that in many cases not the full spectrum of the IMF is covered.}
\end{center}
\end{table}







\end{document}